\documentclass[aps,prb,twocolumn,groupedaddress,showpacs,10pt]{revtex4-1}

\usepackage{graphicx}
\usepackage{amsmath}
\usepackage{amssymb}
\usepackage{color}
\usepackage{hyperref}

\begin{document}

\title{Hall number, specific heat and superfluid density of overdoped high-$T_c$ cuprates}
%\shorttitle{Hall effect and Fermi surface reconstruction in the high-$T_c$ cuprates.} %Insert here a short version of the title if it exceeds 70 characters

\author{J.G. Storey}
%\shortauthor{J.G. Storey}
\affiliation{Robinson Research Institute and MacDiarmid Institute for Advanced Materials and Nanotechnology, Victoria University of Wellington - 
P.O. Box 600, Wellington, New Zealand}
%\institute{                    
%  \inst{1} Robinson Research Institute, Victoria University - 
%P.O. Box 600, Wellington, New Zealand
%}

%\pacs{74.25.F-,74.25.Jb}

\begin{abstract}
%This paper is about the heavily overdoped region where Tc goes to zero. Considered to be less exotic from a physics standpoint, it has received less attention than the rest of the phase diagram and of course in many cuprates this region simply isn’t accessible. % Yet even in this region, the superconducting state is still the location of (is still host to) unexpected/unusual behaviour. Territory?
Despite often being dismissively described as exhibiting conventional Fermi-liquid-like behaviour, heavily overdoped high-$T_c$ cuprates sport several unexpected features. Thermodynamic properties expected to be roughly constant with doping decrease towards zero, signalling that a growing fraction of carriers remain in the normal state below $T_c$. Near $T_c$, the superconducting energy gap fills in with temperature, contrary to the expectations of BCS theory.
Most recently a transition in the Hall number of some cuprates was found to extend to a very high doping ($x\approx0.27$), far beyond the pseudogap critical point identified by a peak in thermodynamic properties ($x$ = 0.19). This presents a challenge to the view that the pseudogap is a consequence of Fermi surface reconstruction. In this paper we present a consistent explanation for all these observations by combining pair-breaking scattering with a Fermi surface reconstruction model for the pseudogap.
Notably, an increase in pair-breaking with doping leads to a separation of the points where reconstruction begins and the thermodynamic properties peak.
This result highlights pair-breaking as an essential ingredient in the electronic recipe for heavily overdoped cuprate superconductors.
\end{abstract}

%In recent years, ARPES and STM experiments have revealed that in cuprates as the temperature is increased the superconducting gap does not shrink but fills-in. Independently, measurements of the magnetic penetration depth, as well as THz spectroscopy, non-linear diamagnetism, specific heat and ultrafast pump-probe measurements all showed that even at the lowest temperatures a fraction of mobile charge carriers remains ‘normal’ and does not join the superfluid. This fraction grows with doping while the low-T superfluid density and Tc decrease, tracking one another.
 
%The above experimental findings are unusual and potentially quite important for ultimate understanding of the high-Tc superconductivity in cuprates. The aim of this session is to present a comprehensive review of the most recent developments in this area and facilitate development of a unified physical picture.

%contrary rather than conflict

\maketitle
\section{Introduction}
The underdoped regime of the cuprate phase diagram is frequently explored because of its rich selection of exotic phenomena, including the pseudogap\cite{Kordyuk2015,Tallon2020}, superconducting fluctuations\cite{WANGNERNST,Alloul_2010} and spin/charge ordering\cite{Tranquada2013,COMIN2016,WEN2019,Frano_2020}. In contrast, the heavily overdoped regime, not easily accessed in many cuprates, has largely been considered the realm of conventional Fermi-liquid-like behaviour\cite{Broun2008170,Keimer2015179}. It is here that we encounter familiar features such as $T$-squared resistivity\cite{MANAKO,NAKAMAE} and a large predictable Fermi surface\cite{SINGH1992193,HUSSEY,PLATEPRL,VIGNOLLE,Rourke_2010}. 
Yet even here, the superconducting phase is host to unusual features which could potentially be important for an ultimate understanding of high-$T_c$ superconductivity.

As doping, $x$, increases towards the edge of the superconducting phase where $T_c(x) \rightarrow 0$:
\textbf{i)} A growing fraction of carriers in the ground state remain normal (i.e. they do not join the condensate). This is evidenced by specific heat\cite{WADE,LORAM2,ENTROPYLSCO,ENTROPYDATA2,WANG}, NMR\cite{OHSUGI}, magnetic penetration depth\cite{Uemura1993605,Niedermayer,BERNHARD1995,LOCQUET,LEMBERGER,Bozovic2016309} and THz spectroscopy\cite{MAHMOOD} measurements. 
\textbf{ii)} The jump in specific heat at $T_c$ and the superfluid density at $T$ = 0  decrease towards zero (a consequence of i)) 
\textbf{iii)} As temperature increases to $T_c$, angle-resolved photoemission spectroscopy (ARPES)\cite{REBER2013,reber2015pairing,KONDO2015}, tunneling\cite{DIPASUPIL2001604,BENSEMAN2018,ZHOU2020} and Raman spectroscopy\cite{GUYARD,GUYARDPRL} experiments have revealed that the superconducting gap fills-in rather than closes. This is at odds with BCS theory\cite{BCSTHEORY}, even though the ratio of the gap magnitude to $T_c$ is close to the $d$-wave weak-coupling BCS value\cite{WANG,He62}.
\textbf{iv)} A recent preprint reports that in some cuprates the Hall number (a measure of the carrier density) only just completes a more gradual transition from $x$ to $1+x$\cite{putzke2019reduced}, challenging a proposition\cite{STOREYHALL} that it is tied to the normal-state pseudogap whose effects typically vanish at a postulated critical point slightly above optimal doping\cite{OURWORK1,Tallon2020}.
%Most recently an increase in Hall number was found to extend to very high doping, far beyond the critical doping traditionally associated with the onset of the pseudogap.

Can these observations be reconciled? i), ii) and iii) can be linked by pair-breaking, but iv) presents a challenge to ii). Namely, if the pseudogap were to close at a higher doping, as suggested by the Hall number, then we would expect the specific heat jump and superfluid density to increase up to that point.
In this paper we will show that the opposite is in fact possible. Our hypothesis is that pair-breaking is the dominant factor in the overdoped regime. We will address the Hall number first before continuing to the specific heat and superfluid density.

%Superfluid density measurements on Tl2201 peak near $x$ = 0.19 consistent with YBCO.

\section{Hall Number}

\begin{figure}
\centering
\includegraphics[width=\linewidth]{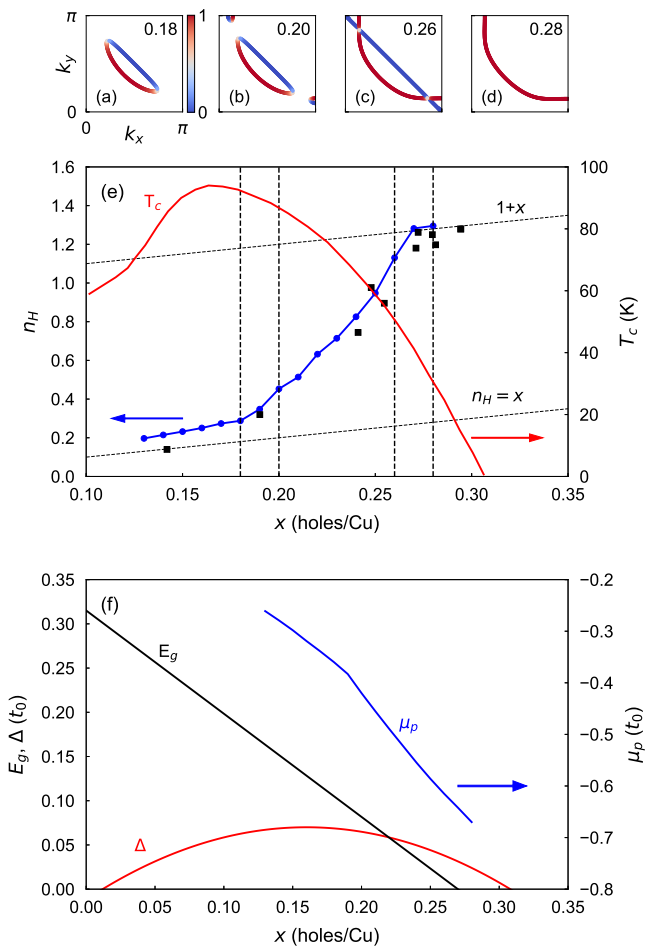}
\caption{
(Color online) (a) to (d) Spectral weight at the Fermi level (red=1, blue=0) calculated from the YRZ model for $x$ = 0.18, 0.20, 0.26 and 0.28. (e) Calculated zero-temperature Hall number vs doping (circles) compared with measured values for Tl2201 (squares) from ref.~\onlinecite{putzke2019reduced}. The measured $T_c$ phase curve from ref.~\onlinecite{putzke2019reduced} is also shown. Vertical lines mark the doping values of panels (a) to (d). (f) Doping dependence of the pseudogap $E_g$, superconducting gap $\Delta$, and chemical potential $\mu_p$ used in the YRZ calculations, tailored to Tl2201. $t_0$ is the unrenormalized nearest-neighbour hopping parameter.
} 
\label{HALLFIG}
\end{figure}

In 2016 the author showed that Fermi surface reconstruction models involving nodal hole-like pockets and antinodal electron-like pockets would result in a transition in the Hall number from $1+x$ to $x$ \cite{STOREYHALL}. One of these models, proposed by Yang, Rice and Zhang (YRZ) \cite{YRZ}, displayed remarkable agreement with measurements on YBa$_2$Cu$_3$O$_{7-\delta}$ (YBCO) \cite{BADOUX}, without any adjustment. Because the onset of this transition coincides with the independently determined pseudogap critical point at $x$ = 0.19\cite{OURWORK1,Tallon2020}, this result supports the view that the pseudogap originates from Fermi surface reconstruction. In YBCO the transition occurs over a narrow doping range $0.19>x>0.16$ According to the models, this is where the electron-like pockets are present on the Fermi surface, though they have not yet been observed directly. %Below $x \approx$  0.165 Hall number is roughly proportional to $x$ which, according to the model, is where the electron-like pockets are absent Fermi surface (they have not yet been observed directly).
%(Do I need to talk about Nd-LSCO? Could mention it here. Has different p* (independently determined?). Perhaps in the conclusion. Change to present tense, e.g. supports)

The present study is motivated in part by recent measurements of the Hall number on Tl$_2$Ba$_2$CuO$_{6+\delta}$ (Tl2201) and Pb/La-doped Bi$_2$Sr$_2$CuO$_{6+\delta}$ (Bi2201) \cite{putzke2019reduced} \footnote{At the time of writing this work is only a preprint, but we assume that the data are reliable.} which show a more gradual transition than YBCO, with the decrease in Hall number beginning at a much higher doping near $x$ = 0.27 and ending near 0.19.
As shown earlier \cite{STOREYHALL}, in the models it is possible to tune onset and slope of the transition by altering the doping dependence of the pseudogap energy parameter $E_g(x)$. So in fig.~\ref{HALLFIG}(e) we plot the Hall number $n_H$ from YRZ using the same approach as before \cite{STOREYHALL}, but with the pseudogap opening below $x$ = 0.27 according to $E_g(x)= 1.167t_0(0.27 - x)$. ($E_g(x)$ and the chemical potential $\mu_p(x)$ are plotted in fig.~\ref{HALLFIG}(f).) This produces a good match with the data (also plotted, sans error bars). For reference, the Fermi surface is plotted for selected doping values showing the evolution with decreasing doping from a large hole-like barrel in fig.~\ref{HALLFIG}(d), to electron- and hole-like pockets in (c) \& (b), and finally to a small hole-like pocket in (a). %In 2013 the author predicted the existence of the electron-like pockets in this doping range Bi$_2$Sr$_2$CuO$_{6+\delta}$ 
The existence of electron-like pockets in this doping range of Bi$_2$Sr$_2$CuO$_{6+\delta}$ was actually predicted by the author in 2013\cite{STOREYTEP}. In that work, it was proposed that electron pockets were responsible for an undulation in the $T$-dependent thermoelectric power seen at $\delta$ = 0.13 and 0.14, corresponding to a doping of $x\sim0.25$\cite{KONSTANTINOVIC}. 
%$E_g(x) = 1.167t_0(0.27 - x)$ for $x\leq0.27$, and zero otherwise.

But there is a problem. In the case of Tl2201 $x$ = 0.27 is far beyond the pseudogap critical point as inferred from measurements of superfluid density ($\approx 0.19$)\cite{Niedermayer}, to be discussed later.  Furthermore, the presence of a pseudogap up to $x$ = 0.27 seems to be at odds with the specific heat jump which decreases with $x$ in that region\cite{WADE,LORAM2,WADETHESIS}, which we will investigate next. Naturally, this raises the question as to whether Fermi surface reconstruction is still an appropriate interpretation of the doping dependence of the Hall number. Note that several alternative explanations for the Hall number have been proposed\cite{EBERLEIN,CHATTERJEE2016,CAPARA,MAHARAJ,VERRET,CHARLEBOIS,SHARMA,MITSCHERLING,BONETTI}. But given that models such as YRZ account for many other observed properties\cite{RICE,JAMES,STOREYEPL2012,ASHBY,STOREYTEP,Storey_2017}, it seems premature to abandon them before considering other contributing factors. One of those factors is pair-breaking.

%(Note that following YRZ we denote doping by $x$, which is often represented by $p$ elsewhere.)

\section{Specific Heat}

%[Talk about: decrease in specific heat jump and increase in $\gamma(0)$ with doping implies an Gamma pair (or Delta/Gamma pair). The steep increase in Gamma pair with temperature near $T_c$ is responsible for the fluctuation tail and non-mean-field-like jump.]
%Note pairbreaking in Zn-doped YBCO looks like Tl2201.

%Thallium 2201 is another system that can be well overdoped, and here the collapse in the jump and the rise in the residual low-temperature term is very obvious. In contrast to LSCO the normal state is pretty flat and doping independent, suggesting that the van Hove singularity is still somewhat below the Fermi level.

\begin{figure}
\includegraphics[width=\linewidth]{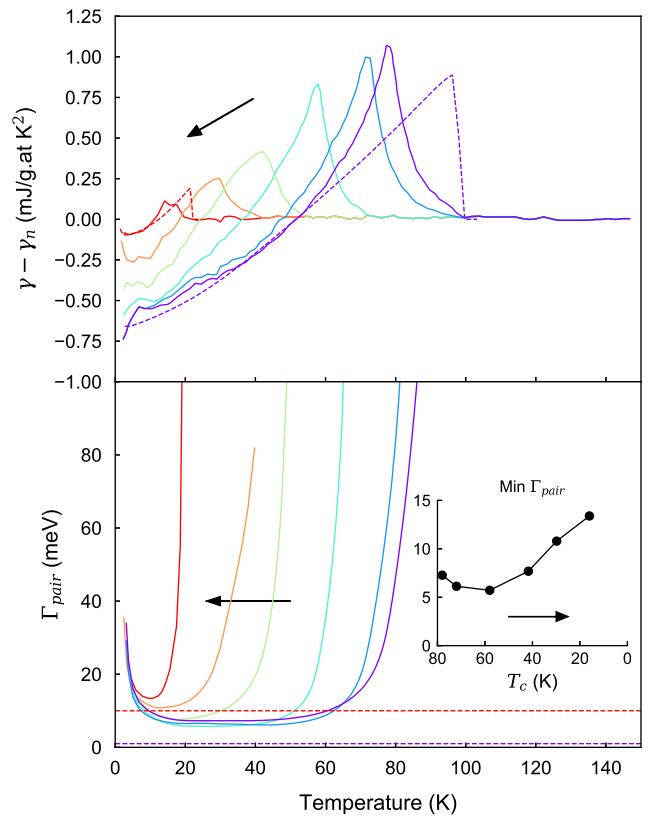}
\caption{
(Color online) (a) Measured specific heat of overdoped Tl2201 with the normal state subtracted (solid lines) from ref.~\onlinecite{WADETHESIS}. Dashed lines show mean-field curves calculated for $\Delta/k_BT_p$ = 2.5 with constant pair-breaking scattering. (b) Pair-breaking scattering rate extracted from the data in (a). Dashed lines correspond to the mean-field curves in (a). The inset shows the minimum value of $\Gamma_{pair}$ vs $T_c$ (estimated by the peak in specific heat). Arrows indicate increasing doping.
} 
\label{WADEFIG}
\end{figure}

In cuprates that can be overdoped to the edge of the superconducting dome the jump in specific heat coefficient at $T_c$, $\delta\gamma(T_c)$, decreases with doping, and the residual term at low temperatures, $\gamma(0)$, increases. Examples include Tl2201\cite{WADE,LORAM2,WADETHESIS} (see fig.~\ref{WADEFIG}(a)) and La$_{2-x}$Sr$_x$CuO$_4$\cite{ENTROPYLSCO,ENTROPYDATA2,WANG}. 
Together these features indicate that some of the carriers are not participating in superconductivity. 
A similar trend is found by NMR where, in the superconducting state, the normalized residual density of states (DOS) at the Fermi level increases remarkably beyond $x\sim0.20$\cite{OHSUGI}. A compilation of these results can be found in fig.~5(b) of ref.~\onlinecite{WANG}.  
(Note that below $x\sim0.19$ the jump size also decreases, but this is accompanied by small residual values\cite{MOMONO1994395}, and implies the existence of a normal-state pseudogap\cite{ENTROPYDATA2}.)

%A similar trend is found by NMR. In the superconducting state, the normalized residual density of states (DOS) at the Fermi level increases remarkably beyond x~0.20. The authors attribute the Tc suppression in overdoped samples to pair-breaking associated with a possible structural inhomogeneity. Doesn’t stack up when we look our next compound.$\gamma(0)$ increases steeply above $x \sim 0.22$, 
%Wang says not due to impurity scattering, suggests phase separation. They attribute this trend to the presence of normal metallic regions, rather than impurity scattering. A possible picture is doped holes going into the Cu3d orbitals, disturbing the antiferromagnetic correlations and forming a normal region.This then leads to a suppression of Tc due to a reduction in pairing strength. 

In this section we will analyse the specific heat of Tl2201 using the two-lifetime phenomenological self-energy proposed by Norman \textit{et al}.\cite{NORMAN}
\begin{equation}
\Sigma(\textbf{k},\omega)=-i\Gamma_{\rm{single}}+\frac{\Delta^2(T)}{\omega+\xi(\textbf{k})+i\Gamma_{\rm{pair}}(T)}
\label{eq:ansatz}
\end{equation}
$\Gamma_{single}$ is a single-particle scattering rate and $\Gamma_{pair}$ is a pair-breaking scattering rate. The method is the same as used previously on optimally-doped Bi$_2$Sr$_2$CaCu$_2$O$_{8+\delta}$ (Bi2212)\cite{Storey_2017_2}. In essence we assume a suitable temperature-dependent superconducting gap $\Delta(T)$, and then find the $\Gamma_{pair}(T)$ that reproduces the observed specific heat. 
Here we take a rescaled $d$-wave BCS gap with magnitude $\Delta_0 = 2.5k_BT_p$ (slightly higher than the weak-coupling value of $2.14k_BT_p$), that opens at the onset of superconducting fluctuations at $T_p>T_c$.
For simplicity we will assume that $\Gamma_{single}$ = 1 meV is negligible compared with $\Gamma_{pair}$.

The specific heat data of overdoped Tl2201 reproduced from ref.~\onlinecite{WADETHESIS} is shown in fig.~\ref{WADEFIG}(a). For comparison dashed curves show the specific heat calculated with a constant $\Gamma_{pair}$ for the highest and lowest doping values. These show a mean-field-like jump extending up to the onset of the fluctuations. The larger is $\Gamma_{pair}$, the greater the residual specific heat.  Figure~\ref{WADEFIG}(b) shows the temperature-dependent $\Gamma_{\rm{pair}}$ curves that reproduce the data in (a). Ignoring the upturns at low-temperature which reflect the small impurity-induced upturns in the specific heat data, there does seem to be an increase in the base value of $\Gamma_{pair}$ with doping. The inset to (b) shows that the minimum value increases as $T_c$ (estimated from the peak in specific heat) goes to zero. As found previously\cite{Storey_2017_2}, near $T_c$ the pair-breaking scattering rate diverges, reflecting superconducting fluctuations, and is responsible for the gap-filling behaviour observed in ARPES\cite{REBER2013,reber2015pairing,KONDO2015}, tunneling\cite{DIPASUPIL2001604,BENSEMAN2018,ZHOU2020} and Raman\cite{GUYARD,GUYARDPRL} spectroscopies.

So a $T$- and $x$-dependent $\Gamma_{pair}$ accounts for points i), ii) and iii) in the introduction. Can the pseudogap open under these circumstances, thereby reconciling point iv)?
To answer this question we incorporate the self energy (\ref{eq:ansatz}) into the YRZ Green's function as follows
\begin{equation}
G(\textbf{k},\omega)=\sum_{\alpha=\pm}{\frac{W_\textbf{k}^\alpha}{\displaystyle \omega-E_\textbf{k}^\alpha+i\Gamma_{\rm{single}}-\frac{\Delta_{\textbf{k}}^2(T)}{\omega+E_\textbf{k}^\alpha+i\Gamma_{\rm{pair}}}}}
\label{eq:GYRZ2}
\end{equation}
where, following ref.~\onlinecite{STOREYHALL}, we have dropped the $x$-dependent Gutzwiller prefactor. 
$E_\textbf{k}^\pm$ are the upper and lower branches of the YRZ-reconstructed dispersion,
\begin{equation}
E_\textbf{k}^\pm=\frac{\xi_\textbf{k}-\xi_\textbf{k}^0}{2}\pm\sqrt{\left(\frac{\xi_\textbf{k}+\xi_\textbf{k}^0}{2}\right)^2+E_g^2(\textbf{k})}
\label{eq:EK}
\end{equation}
and $W_\textbf{k}^\pm$ are corresponding the weight factors
\begin{equation}
W_\textbf{k}^\pm=\frac{1}{2}\left[1\pm\frac{(\xi_\textbf{k}+\xi_\textbf{k}^0)/2}{\sqrt{[(\xi_\textbf{k}+\xi_\textbf{k}^0)/2]^2+E_g^2(\textbf{k})}}\right]
\label{eq:WK}
\end{equation}
$\xi_\textbf{k}=-2t(x)(\cos k_x+\cos k_y)-4t^\prime(x)\cos k_x\cos k_y-2t^{\prime\prime}(x)(\cos 2k_x+\cos 2k_y)-\mu_p(x)$ is the tight-binding energy-momentum dispersion and $\xi_\textbf{k}^0=-2t(x)(\cos k_x+\cos k_y )$ is the nearest-neighbour term. The tight-binding coefficients are the same as before\cite{STOREYHALL}. The pseudogap is $E_g(\textbf{k})=E_g(x)(\cos(k_x)-\cos(k_y))/2$ and the superconducting gap is $\Delta_{\textbf{k}}(T)=\Delta(x)\delta(T)(\cos(k_x)-\cos(k_y))/2$. $\delta(T)$ is the normalised $d$-wave BCS temperature dependence.
The pseudogap magnitude $E_g(x)$ and chemical potential $\mu_p(x)$ are taken from the Hall number calculations in the previous section. We also adopt a superconducting gap given by $\Delta(x)=0.07t_0[1.0 - 45.6(x - 0.16)^2]$ (shown in fig.~\ref{HALLFIG}(f)), to match the wider doping range spanned by the observed Tl2201 $T_c$ dome\cite{putzke2019reduced,BANGURA2010}. 
The density of states is calculated from
\begin{equation}
N(\omega)=\sum_{\textbf{k}}{A(\textbf{k},\omega)}
\label{eq:DOS}
\end{equation}
where the spectral function $A(\textbf{k},\omega)=\pi^{-1}\rm{Im}G(\textbf{k},\omega)$. 
Finally, the electronic specific heat coefficient $\gamma(T)$ is calculated in the usual way from 
\begin{equation}
\gamma(T) = -2k_{\rm B}\frac{\partial}{\partial T}\int{[f\ln f + (1-f)\ln (1-f)]N(\omega)d\omega}
\label{eq:gamma}
\end{equation}
where $f$ is the Fermi-Dirac distribution.

\begin{figure}
\centering
\includegraphics[width=\linewidth]{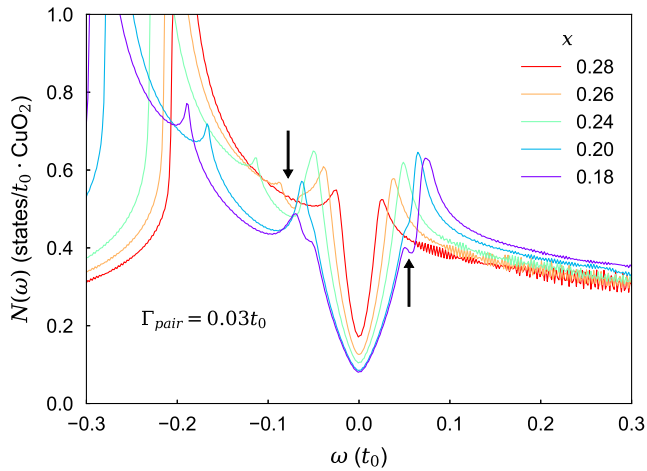}
\caption{
(Color online) Calculated density of states per spin at $T$ = 0 with pair-breaking scattering $\Gamma_0$ = 0.03$t_0$ for several values of $x$. The downwards arrow indicates the pseudogap in the $x$ = 0.26 curve caused by the onset of Fermi surface reconstruction. As doping decreases the pseudogap spans the Fermi level which leads to sub-gap structure indicated by the upwards arrow.
} 
\label{DOSFIG}
\end{figure}

The density of states is plotted in fig.~\ref{DOSFIG} for several values of $x$ with $\Gamma_{pair}$ = 0.03$t_0$. (For simplicity $\Gamma_{single}$ is set to a negligible value, = 0.001$t_0$) The pseudogap first appears as a depression at negative energies below the superconducting coherence peak, and doesn't impact states in the immediate vicinity of the Fermi level. As a result, the decrease in $\delta\gamma(T_c$) (with decreasing $x$) is not locked to the onset of Fermi surface reconstruction\cite{VERRET}. As doping reduces, the depression deepens and eventually spans the Fermi level. This happens just below $x$ = 0.2 and coincides with the lifting of the antinodal electron-like pockets from the Fermi surface, as seen in figs.~\ref{HALLFIG}(a) \& (b), and the appearance of sub-gap structure\cite{MCELROY} in the density of states (see fig.~\ref{DOSFIG}). 

\begin{figure}
\centering
\includegraphics[width=\linewidth]{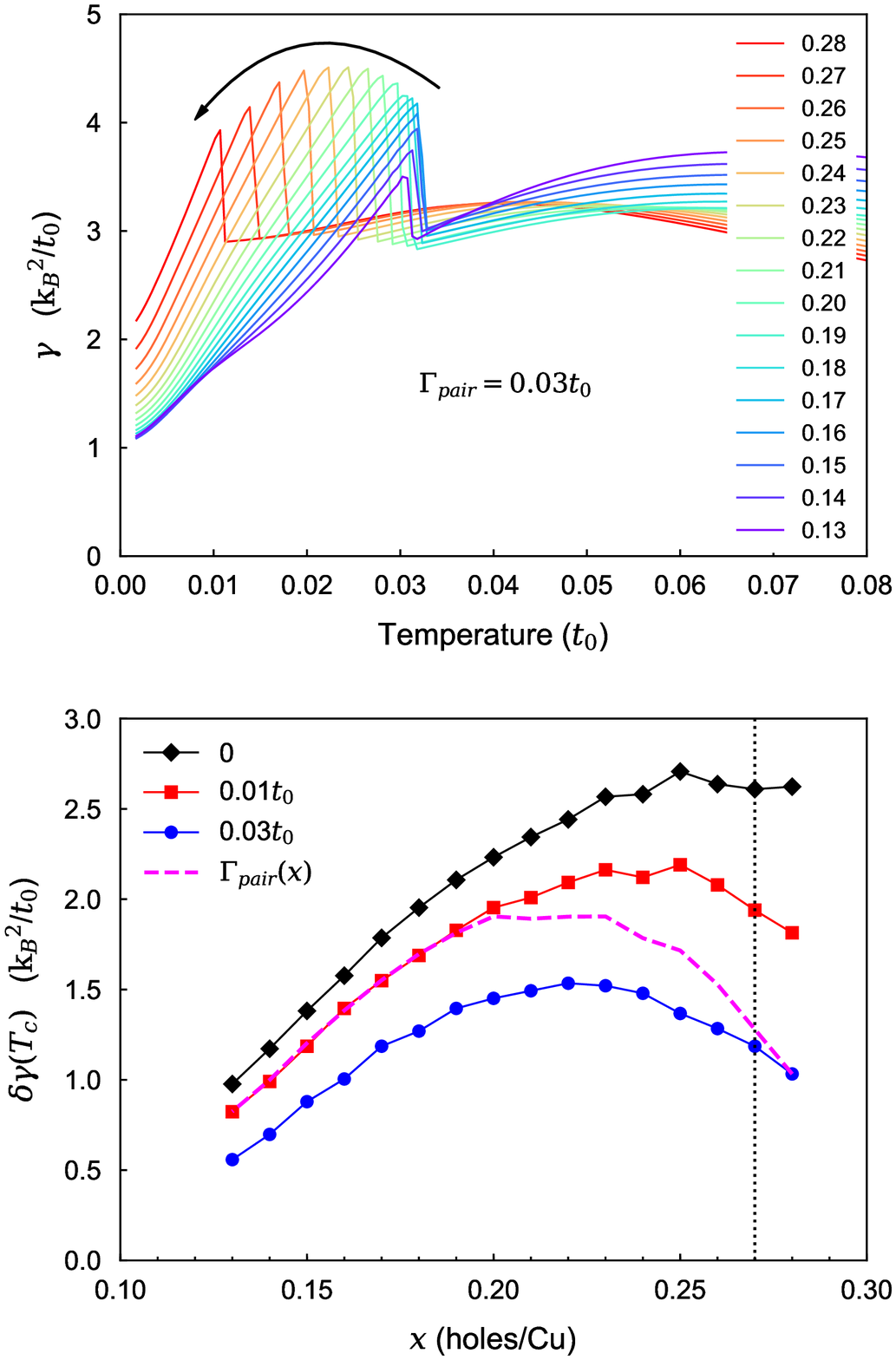}
\caption{
{(Color online) Specific heat coefficient calculated from the YRZ model using the parameters in fig.~\ref{HALLFIG}(f) with pair-breaking scattering $\Gamma_{pair}$ = 0.03$t_0$. The arrow denotes increasing doping. (b) Doping dependence of the specific heat jump at $T_c$ for $\Gamma_{pair}$ set to 0, 0.01$t_0$, 0.03$t_0$ and $\Gamma_{pair}(x)$ as described in the text. The vertical dotted line marks the onset of Fermi surface reconstruction.}
} 
\label{GAMMAFIG}
\end{figure}

The resulting temperature-dependent specific heat coefficient is plotted in fig.~\ref{GAMMAFIG}(a) for $x$ ranging from 0.28 to 0.13. With increasing doping, the residual low-$T$ value increases due to the decreasing $\Delta/\Gamma_{pair}$ ratio. Meanwhile the jump at $T_c$  increases up to $x$ = 0.23, before decreasing due to pair-breaking dominating over the closing pseudogap. A more detailed treatment could include a $T$-dependent $\Gamma_{pair}$ to reproduce the tails on the jump from fluctuations. As shown in fig.~\ref{GAMMAFIG}(b), the peak in $\delta\gamma(T_c)$ can be tuned by the value of $\Gamma_{pair}$. For example, if $\Gamma_{pair}(x)$ increases linearly from 0.01$t_0$ at $x$ = 0.19 to 0.03$t_0$ at $x$ = 0.28, then $\delta\gamma(T_c)$ traces the path of the dashed line, which peaks nearer to $x$ = 0.19. We note that a peak in $\delta\gamma(T_c)$ has not yet been observed in Tl2201 (as it difficult to underdope), but it has in Bi2201\cite{WEN2009}.  

So to answer our question, the addition of pair-breaking makes it possible for the pseudogap to open, while the specific heat jump at $T_c$ increases. Remember that the Hall number is a normal-state property and is unaffected by $\Gamma_{pair}$.
Next we consider the superfluid density.

\section{Superfluid Density}

\begin{figure}
\includegraphics[width=\linewidth]{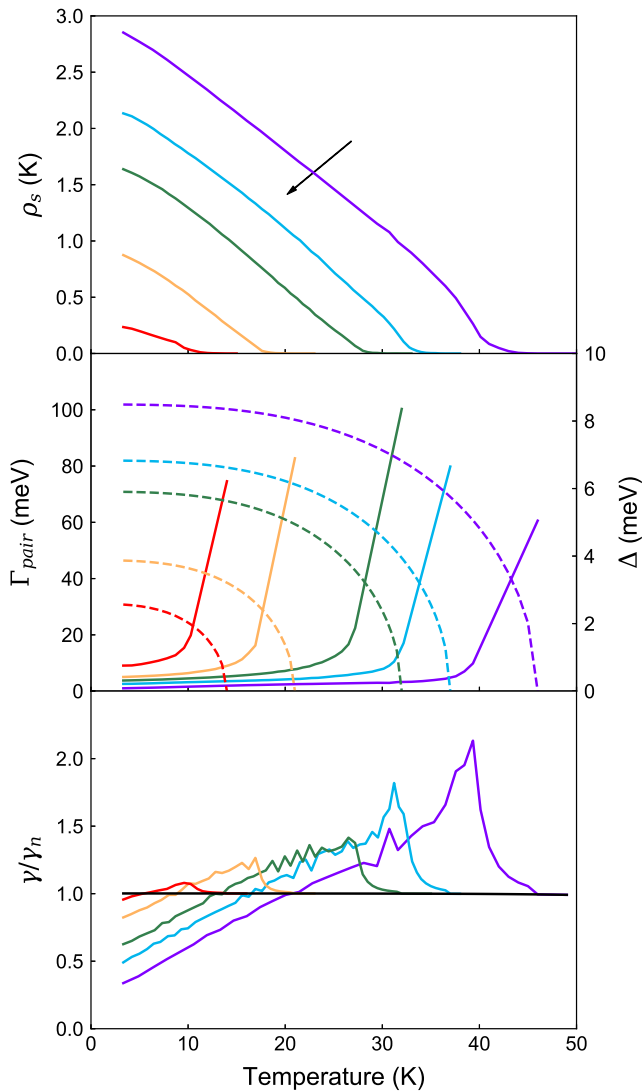}
\caption{
(Color online) (a) Superfluid density measured on overdoped LSCO from ref.~\cite{Bozovic2016309} The arrow indicates the direction of increasing doping. (b) Assumed BCS superconducting gaps (dashed), and pair-breaking scattering rates (solid) extracted from the data in (a). (c) Specific heat coefficient calculated using the parameters in (b). %(subtract normal state so can better compare with Tl2201? Mention that normal state is not flat in LSCO) (Do we need another inset or figure showing the doping dependence of Gamma0?) 
} 
\label{LSCOFIG}
\end{figure}

In the absence of the pseudogap and pair-breaking we expect the superfluid density (proportional to the inverse square of the penetration depth) to scale with the normal-state carrier density. But according to muon-spin rotation experiments on Tl2201\cite{Uemura1993605,Niedermayer}, Y$_{0.8}$Ca$_{0.2}$Ba$_{2}$(Cu$_{1-z}$Zn$_z$)$_3$O$_{7-\delta}$ and Tl$_{0.5-y}$Pb$_{0.5+y}$Sr$_{2}$Ca$_{1-x}$Y$_{x}$Cu$_{2}$O$_{7}$\cite{BERNHARD}, the zero-temperature superfluid density, $\rho_s(0)$, peaks near $x$ = 0.19 before decreasing with further overdoping. The resulting plot of $T_c$ vs $\rho_s(0)$ follows a so-called ``boomerang'' trajectory\cite{BERNHARD1995} back to the origin. The anomalous scaling of $\rho_s(0)$ with $T_c$ and its eventual decrease to zero with overdoping has also been studied in great detail by mutual inductance experiments on LSCO thin films\cite{LOCQUET,LEMBERGER,Bozovic2016309}. It can be shown that these results are congruent with the boomerang plot\footnote{J.L. Tallon, private communication}.

%It also reveals an unexpected linear temperature dependence. Jeff (private comm) has recently compiled an updated boomerang plot. Falls on the same line. Tl2201 musr (and specific heat).

%Increase up to 0.19 was presumed to be due to the closing pseudogap.
%But instead it bends back around and heads back linearly to the origin. The so-called boomerang shape. 

%Puźniak et al., PRB 53, 86 (1996) A decrease in superfluid density is also found in mercury 1201 with no increase in the effective mass.  This points to coexistence of paired and unpaired fermions in the overdoped region of high-Tc superconductors at T=0.

%And just to show you that this isn’t some artefact of muon spin rotation, the photoemission peak intensity in the superconducting state following the same trend. Feng et al., Science 289, 277 (2000)

%Lee-Hone et al., PRB 96, 024501 (2017) The data is replotted here in this nice paper where the authors have done a good job reproducing the behaviour with a dirty d-wave model incorporating a blend of weak and strong scattering.

Like the specific heat, we will analyse the superfluid density of LSCO using the two-lifetime model (\ref{eq:ansatz}). Again, the approach is the same as used previously for optimally-doped Bi2212\cite{Storey_2017_2}, and involves making a choice for the $\Delta(T)$ and then finding the $\Gamma_{pair}(T)$ that reproduces the data. A selection of $T$-dependent superfluid density curves for overdoped LSCO\cite{Bozovic2016309} (digitized from ref.~\onlinecite{LEEHONE}) are shown in fig.~\ref{LSCOFIG}(a). They are approximately linear in $T$  and the low-temperature values scale roughly with $T_c$.
A $d$-wave BCS gap that opens at the onset of superconducting fluctuations was assumed for each curve, see fig.~\ref{LSCOFIG}(b), and the extracted $\Gamma_{pair}(T)$ curves are shown in the same panel. These look similar to $\Gamma_{pair}(T)$ extracted from the specific heat data in fig.~\ref{WADEFIG}(b). There is an increase in low-$T$ values with doping and a divergence near $T_c$ from fluctuations. Figure~\ref{LSCOFIG}(c) shows the specific heat coefficient calculated from the parameters in panel (b). The curves are a little noisy but display the same features as the Tl2201 data, namely an increasing $\gamma(0)$ and decreasing $\delta\gamma(T_c)$. This demonstrates that the LSCO and Tl2201 samples share a common phenomenology despite their significant differences in crystal structure and morphology.

Following the same pattern as the previous section, we will now see if it is possible for the pseudogap to open below $x$ = 0.27 while $\rho_s(0)$ increases.
We insert the self energy (\ref{eq:ansatz}) into the YRZ anomalous Green's function $F$ as follows
\begin{equation}
\begin{split}
F(\textbf{k},\omega)&=F^+(\textbf{k},\omega)+F^-(\textbf{k},\omega)
\\&=\sum_{\alpha=\pm}\frac{W_\textbf{k}^\alpha\Delta_{\textbf{k}}(T)/(\omega+E_\textbf{k}^\alpha+i\Gamma_{\rm{pair}})}{\displaystyle\omega-E_\textbf{k}^\alpha+i\Gamma_{\rm{single}}-\frac{\Delta^2_{\textbf{k}}(T)}{\omega+E_\textbf{k}^\alpha+i\Gamma_{\rm{pair}}}}
\end{split}
\label{eq:F}
\end{equation}
Again for simplicity $\Gamma_{single}$ is set to a negligible value, = 0.001$t_0$.
Generalizing the expression for the superfluid density\cite{CARBOTTE} for the two branches of the YRZ dispersion, we have
\begin{equation}
\begin{split}
\frac{1}{\lambda^2(T)} &= \frac{16\pi e^2}{c^2V}\sum_\textbf{k}\int{d\omega^\prime d\omega^{\prime\prime}\lim_{q \rightarrow 0}\left[\frac{f(\omega^{\prime\prime})-f(\omega^\prime)}{\omega^{\prime\prime}-\omega^\prime}\right]}\\&\times [v^+_xB^+(\textbf{k+q},\omega^\prime)+v^-_xB^-(\textbf{k+q},\omega^\prime)]\\&\times [v^+_xB^+(\textbf{k},\omega^{\prime\prime})+v^-_xB^-(\textbf{k},\omega^{\prime\prime})]
\end{split}
\label{eq:lsq1}
\end{equation}
where $B^\pm(\textbf{k},\omega)=\pi^{-1}\rm{Im}F^\pm(\textbf{k},\omega)$ and the group velocities of each branch are $v^\pm_x=\partial E^\pm_{\textbf{k}}/\partial k_x$.

\begin{figure}
\centering
\includegraphics[width=\linewidth]{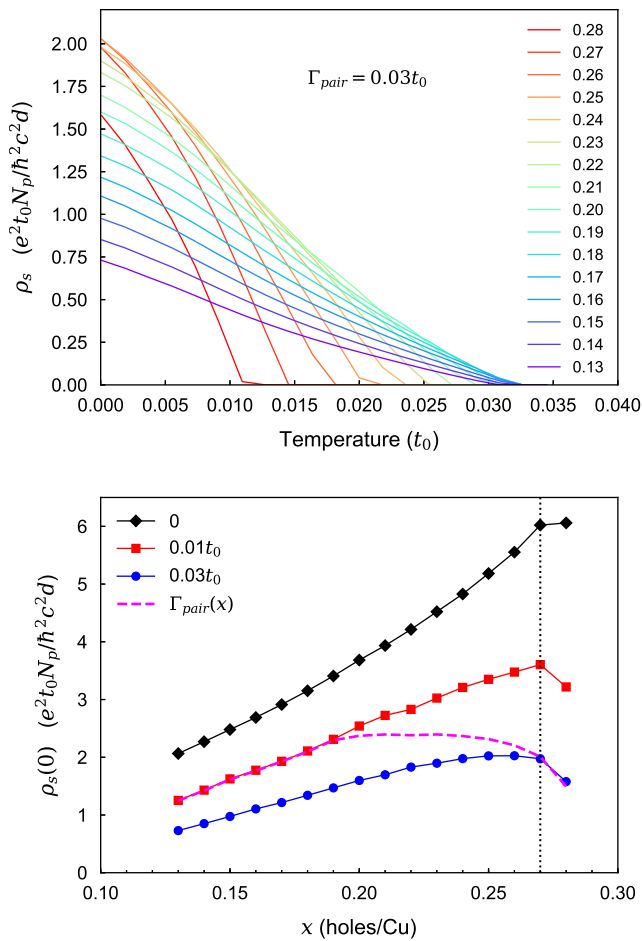}
\caption{
{(Color online) (a) Superfluid density calculated from the YRZ model using the parameters in fig.~\ref{HALLFIG}(f) with pair-breaking scattering $\Gamma_{pair}$ = 0.03$t_0$. (b) Doping dependence of the superfluid density at $T$ = 0 for $\Gamma_{pair}$ set to 0, 0.01$t_0$, 0.03$t_0$ and $\Gamma_{pair}(x)$. The vertical dotted line marks the onset of Fermi surface reconstruction. The units are ${e^2t_0N_p}/{\hbar^2 c^2 d}$ where $N_p$ is the number of CuO$_2$ planes per unit cell, and $d$ is the $c$-axis lattice parameter.}
} 
\label{SFDFIG}
\end{figure}

Figure \ref{SFDFIG}(a) shows the temperature-dependent superfluid density $\rho_s(T)$ for $\Gamma_{pair}$ = 0.03$t_0$ and $x$ ranging from 0.28 to 0.13. With decreasing doping, $\rho_s(0)$ increases up to about $x\sim0.255$ before succumbing to the pseudogap and decreasing. In contrast to the specific heat jump, the peak in $\rho_s(0)$ is barely shifted from $x$ = 0.27 where the pseudogap opens.
The effect of $\Gamma_{pair}$ is shown in fig.~\ref{SFDFIG}(b). In order for $\rho_s(0)$ to peak near $x$ = 0.19, $\Gamma_{pair}$ would need to decrease down to $x$ = 0.19. This is illustrated by the dashed line which shows the effect of $\Gamma_{pair}$ decreasing linearly from 0.03$t_0$ at $x$ = 0.28 to 0.01$t_0$ at $x$ = 0.19. %Comment of linearity of rhos(T) and $\Gamma_{pair}(T)$?) 

\section{Discussion}
In this section we discuss the findings and pose some questions for future investigation.
We have analysed the specific heat and superfluid density data of overdoped cuprates using an ansatz for the self energy, focussing on the effect of the pair-breaking scattering rate $\Gamma_{pair}$. From this we have deduced that $\Gamma_{pair}$ increases with doping beyond $x\sim0.2$. This is responsible for the decrease in zero-$T$ superfluid density and specific heat jump at $T_c$, as well as the increase in residual specific heat. The divergence of $\Gamma_{pair}$ near $T_c$ reproduces the smooth tails associated with fluctuations. What is the origin of $\Gamma_{pair}$? Assuming that pair-breaking at $T$ = 0 and near $T_c$ share a common origin, then the interactions that give rise to pair-breaking become gapped below the pairing temperature $T_p$. At low temperature, the interactions become less gapped in the strongly overdoped regime, which coincides with a weakening of the pairing strength and reduction in superconducting gap magnitude. A possible candidate for these interactions is the spin fluctuation spectrum. Inelastic neutron scattering studies show a suppression of antiferromagnetic spectral weight with overdoping\cite{WAKIMOTO2004,WAKIMOTO2007}. How can we further understand the pair-breaking? One approach is to look for a momentum dependence. An ARPES study on Tl2201 reported increasing linewidth broadening (a measure of the scattering rate) near the nodal regions of the Fermi surface with doping \cite{PEETS}. 
%which increases on overdoping when going from a sample with $T_c$ = 63 K to a sample with $T_c$ = 30 K\cite{PEETS}. 
Broadening around the nodes is consistent with quantum critical fluctuations. Competing ferromagnetic fluctuations have been proposed\cite{Kopp6123,Maier202015} and experimental evidence for\cite{KURASHIMA,PENG} and against\cite{Wu2017} such fluctuations has been reported.
It is important to note that an alternative description of overdoped cuprates based dirty on $d$-wave BCS theory has been proposed\cite{LEEHONE,LEEHONE2020}, though it has faced some rebuttal by others\cite{Wu2017,BOZOVIC2018}.

By inserting $\Gamma_{pair}$ into the YRZ model we have shown that it is possible for Fermi surface reconstruction to begin far beyond the doping at which the thermodynamic properties peak. This reconciles the seemingly conflicting observations in Tl2201 of a decreasing Hall number, below $x$ = 0.27, and an increasing $\delta\gamma(T_c)$ \& $\rho_s(0)$. At the onset of the reconstruction the pseudogap only affects a small number of occupied states below the superconducting gap edge, and so the thermodynamic properties are governed by pair-breaking. Near 0.19 where $\Gamma_{pair}$ is small, the growing pseudogap spans the Fermi level, becoming the dominant influence and weakening superconducting properties. It is here that the antinodal electron-like pockets lift away from the Fermi surface. Could $\Gamma_{pair}$ be linked to these pockets? Though perhaps coincidental, it's interesting to note that residual specific heat increases near 1/8th doping\cite{MOMONO1994395,TALLON6} where the presence of other electron-like pockets has also been inferred\cite{LeBoeuf2007533}. But there are problems with this idea. Firstly an increase in broadening near the antinodes disagrees with the observed nodal broadening mentioned above. Secondly, the doping range over which the electron pockets are present seems to be material-dependent and, as in the case of YBCO\cite{BADOUX}, can be quite narrow.

Observing the electron pockets directly will be difficult because only one side has appreciable spectral weight. Figure~\ref{HALLFIG}(c) shows that the Fermi surface for $x$ = 0.26 could easily be mistaken for an unreconstructed Fermi surface. The distinguishing feature is a small region of reduced spectral weight slightly inboard of the zone boundaries (where the red and blue lines appear to cross). It has been noted by the authors of ref.~\onlinecite{DAS} that such features are visible in ARPES measurements on overdoped Tl2201\cite{PLATEPRL}. They also calculated the expected quasiparticle-interference maps, however experimentally measured maps\cite{He608} are not clear enough to draw a firm conclusion on the presence of the pockets.

What triggers the Fermi surface reconstruction? The onset doesn't appear to be universal. In YBCO it is near 0.19\cite{BADOUX}, 0.23 in La$_{1.6-x}$Nd$_{0.4}$Sr$_x$CuO$_4$\cite{COLLIGNON} and 0.27 in Tl2201 \& Bi2201\cite{putzke2019reduced}. It has been noted that a hole-like (unreconstructed) Fermi surface seems to be required\cite{Doiron-Leyraud2017}. In other words, the Fermi level must be above the saddle-point van-Hove singularity. Perhaps a self-consistent treatment can identify a connection with band-structure parameters as was done in the case of $s$-wave superconductivity with charge-density-waves\cite{BALSEIRO}. A self-consistent formalism for the two gaps in the YRZ model does exist\cite{Rao_2017}. But we also note that, although we have employed the YRZ model here, similar results can be expected from the antiferromagnetic $Q=(\pi,\pi)$ zone-folding reconstruction model\cite{CHUBUKOV,STOREYHALL}.

Looking across the phase diagram, the following summary can be made. In the underdoped regime there is a loss of normal states. The superconducting gap is large but it is truncated by a small Fermi surface. Pair-breaking is small. In the overdoped regime some states are normal below $T_c$. The Fermi surface is large but the superconducting gap is small. Pair-breaking here plays a significant role.

\begin{acknowledgments}
Supported by the Marsden Fund Council from Government funding, administered by the Royal Society of New Zealand. The author acknowledges helpful discussions with J.L. Tallon.
\end{acknowledgments}

\end{document}